\documentclass{aa}
\usepackage{graphicx}
\usepackage{psfig}
\usepackage{natbib}
\bibpunct{(}{)}{;}{a}{}{,}

\begin{document}

% our macros
\def\spose#1{\hbox to 0pt{#1\hss}}
\def\lta{\mathrel{\spose{\lower 3pt\hbox{$\mathchar"218$}}
     \raise 2.0pt\hbox{$\mathchar"13C$}}}
\def\gta{\mathrel{\spose{\lower 3pt\hbox{$\mathchar"218$}}
     \raise 2.0pt\hbox{$\mathchar"13E$}}}
\def\simless{\mathbin{\lower 3pt\hbox
     {$\rlap{\raise 5pt\hbox{$\char'074$}}\mathchar"7218$}}}   %< or of order
\def\simmore{\mathbin{\lower 3pt\hbox
     {$\rlap{\raise 5pt\hbox{$\char'076$}}\mathchar"7218$}}}   %> or of order
\def\Msun{{\rm M}_\odot}                                       % solar masses
\def\msun{{\rm M}_\odot}                                       % solar masses
\def\Rsun{{\rm R}_\odot}
\def\Lsun{{\rm L}_\odot}
\def\half{{1\over2}}
\def\RL{R_{\rm L}}
\def\zs{\zeta_{s}}
\def\zR{\zeta_{\rm R}}
\def\dJJ{{\dot J\over J}}
\def\dMM{{\dot M_2\over M_2}}
\def\tKH{t_{\rm KH}}
\def\eck#1{\left\lbrack #1 \right\rbrack}
\def\rund#1{\left( #1 \right)}
\def\wave#1{\left\lbrace #1 \right\rbrace}
\def\dd{{\rm d}}
\def\rem#1{{ #1}}
\def\new#1{{ #1}}

\title{The distribution of kHz QPO frequencies in bright LMXBs}

\titlerunning{kHz QPO frequencies in bright LMXBs}

\author{T. Belloni\inst{1}, M. M\'endez\inst{2}
    \and
    J. Homan\inst{1,3}
}

\offprints{T. Belloni}

\institute{INAF -- Osservatorio Astronomico di Brera,
        Via E. Bianchi 46, I-23807 Merate (LC), Italy
   \and
    SRON, National Institute for Space Research, Sorbonnelaan 2,
    3584 CA Utrecht, the Netherlands
  \and
        Center for Space Research, Massachusetts Institute of Technology,
        77 Massachusetts Avenue, Cambridge, MA 02139-4307, USA
}

\date{Received 31 May 2004; accepted 8 January 2005}

\abstract{ 
We analyzed all published frequencies, $\nu_1$ and $\nu_2$,
of the twin kilohertz quasi-periodic oscillations (kHz QPOs) in bright
neutron star low-mass X-ray binaries. The two frequencies are well
correlated but, contrary to recent suggestions, the frequency-frequency
correlation is significantly different from a $\nu_2 = (3/2) \nu_1$
relation.
To check whether, although not following the the 3/2 relation, the QPO
frequencies cluster around a region where $\nu_2/\nu_1 \approx 3/2$,
we re-analyzed the \object{Sco X-1} data that were used to report
that ratio 
\new{and show that,
because the distribution of ratios of linearly correlated measurements is
intrinsically
biased, although the significance of the clustering around $\nu_2/\nu_1
\approx 3/2$ previously reported in the case of \object{Sco X-1} is formally
correct, it does not provide any useful information
about a possible underlying resonance mechanism in this source.} Using
the same data, we then show that the (unbiased) distribution of QPO
frequencies is consistent with a uniform distribution at a $2.4\sigma$ level.
To investigate this further, we analyzed a larger data set of
\object{Sco X-1} and four other sources, 
\object{4U 1608--52}, \object{4U 1636--53}, \object{4U 1728--34}
and 4U 1820--30. We find that for all five sources the distribution of
the kHz QPO frequencies is not uniform and has multiple peaks, which
have no analogy in the distribution of
points in the spectral color-color diagrams of these sources.
\new{Finally, we demonstrate that a simple random walk of the QPO frequencies can
reproduce qualitatively the observed distributions in frequency and
frequency ratio.  This result
weakens the support for resonance models of kHz QPOs in neutron stars.}

}

\maketitle

\section{Introduction}

The launch of the Rossi X-Ray Timing Explorer (RXTE) satellite has led
to the discovery of high-frequency quasi-periodic oscillations (QPOs)
in bright low mass X-ray binaries (LMXBs) containing neutron stars (see
van der Klis 2000 for a review). These QPOs provide a probe into the
accretion flow around neutron stars very close to the compact object,
where effects of general relativity might be observable. They often
appear in pairs at frequencies of a few hundred Hz to more than 1 kHz,
from which the name {\it kHz QPOs} was derived.  Their frequencies
follow rather tight correlations with other timing features of the
X-ray emission (see Ford \& van der Klis 1998; Psaltis et al.
1999; Belloni et al. 2002).

There is currently no consensus as to the origin of these QPOs, nor on
what physical parameters determine their frequencies, which have been
identified with various characteristic frequencies in the inner
accretion flow (see e.g. Stella \& Vietri 1999; Osherovich \& Titarchuk
1999; Lamb \& Miller 2003).

Recently, pairs of 30--450 Hz QPOs have been observed from a few
black-hole candidates and their frequencies tend to appear in certain
ratios (3:2, 5:3, see e.g. Strohmayer 2001a,b; Miller et al. 2001), 
although there are exceptions (see e.g. Homan et al. 2001).
Unfortunately, the detection of such QPOs in black-hole candidates is
rather rare, which makes it difficult to assess the significance of
this clustering around definite ratios. Abramowicz et al. (2003)
reported that the ratio of the frequencies in the kHz QPOs from the
brightest LMXB in the sky, \object{Sco X-1}, tend to cluster around a value of
1.5, which they interpret as evidence for a 3:2 resonance. A
mathematical approach to the resonance model was presented by Rebusco
(2004) in order to explain the \object{Sco X-1} results; the discrepancies of
the data with a pure 3:2 ratio were attributed to the action of an
additional  ad-hoc force.

In this paper, we present the results of an analysis of all published
values of  kHz QPO frequencies from neutron-star systems, including
those from Sco X-1. We also extracted and analyzed long series of
QPO frequencies from a few selected systems:
\object{Sco X-1}, \object{4U 1608--52}, \object{4U 1636--53}, 
\object{4U 1728--34} and \object{4U 1820--30}. From these
data, we conclude that the presence of a single fixed ratio between the
frequencies of these QPOs can be excluded. We do, however, find
evidence of a clustering of QPO frequencies around specific values that
might be associated with  resonances. \new{Additional independent observations
are needed to exclude a random origin of this clustering.}

\section{The sample of published kHz QPO frequencies}

We searched the literature for all published instances of kHz QPO
frequencies. In some cases, no tables are provided and the numbers could
only be obtained from figures. We did extract numbers from such figures
only when no other frequencies were available from that source. We
restricted our sample to the detection of {\it two} simultaneous  kHz
QPO peaks and discarded all data obtained via a shift-and-add technique
(see M\'endez et al. 1998a), \new{as this procedure does not allow to
reconstruct the distribution of pairs of frequencies (see Abramowicz
et al. 2003).} 
The adopted sources with the range of lower
kHz QPO frequency and references are reported in Table 1. For \object{Sco X-1},
we used the same data as in Abramowicz at al. (2003).

\begin{table}
\begin{center}
\caption{\small List of sources considered in this work, with range of
observed $\nu_1$ frequency and references}
\begin{tabular}{lcl}
\hline
\hline
Source   &  Frequency & References \\
 name    &  range (Hz)&            \\
\hline
\multicolumn{3}{c}{Z sources}\\
\hline
{\rm \object{GX 17+2}}     & 475-830 & [1] \\
{\rm \object{Sco X-1}}     & 565-829 & [2]\\
{\rm \object{GX 5-1}}      & 156-627 & [3]\\
{\rm \object{GX 340+0}}    & 197-565 & [4]\\
{\rm \object{Cyg X-2}}     & 532     & [5]\\
{\rm \object{GX 349-2}}    & 712-715 & [6,7]\\
\hline
\multicolumn{3}{c}{Atoll sources}\\
\hline
{\rm \object{4U 1728-34}}  & 308-876 & [8,9]\\
{\rm \object{4U 0614+09}}  & 153-823 & [10]\\
{\rm \object{4U 1705-44}}  & 776     & [11]\\
{\rm \object{KS 1731-260}} & 903     & [12]\\
{\rm \object{4U 1735-44}}  & 641-726 & [13]\\
{\rm \object{4U 1608-52}}  & 550     & [14]\\
{\rm \object{4U 1636-53}}  & 865-920 & [15,16]\\
{\rm \object{4U 1820-30}}  & 790     & [17]\\
{\rm \object{4U 1915-05}}  & 224-707 & [18]\\
{\rm \object{XTE J2123-058}}& 849-871 & [19]\\

\hline
\end{tabular}
\end{center}
{\footnotesize [1] Homan et al. (2002); [2] van der Klis et al. (1997);
[3] Jonker et al. (2002a); [4] Jonker et al. (2000); [5]
Wijnands et al. (1998); [6] Zhang et al. (1998); [7] O'Neill et
al. (2002); [8] Migliari et al. (2003); [9] Di Salvo et al.
(2001); [10] van Straaten et al. (2000); [11] Ford et al. (1998a);
[12] Wijnands \& van der Klis (1997); [13] Ford et al. (1998b);
[14] M\'endez et al. (1998b); [15] Wijnands et al. (1997); [16] Di
Salvo et al. (2003); [17] Smale et al. (1997); [18] Boirin et al.
(2000); [19] Homan et al. (1999).}

\end{table}

\begin{figure}[h] \resizebox{\hsize}{!}{\includegraphics{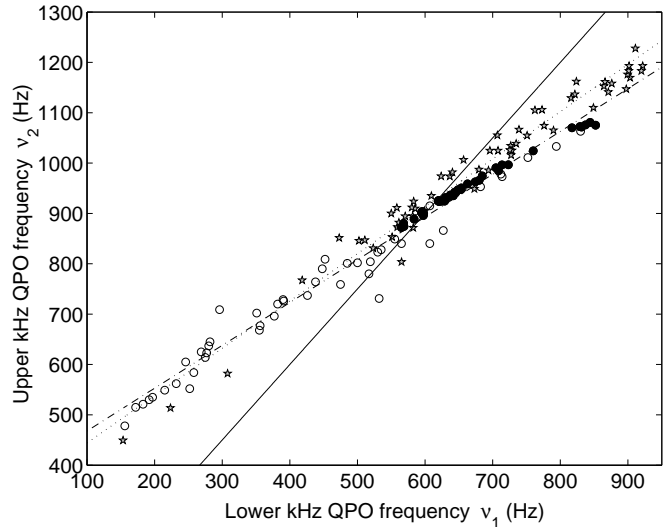}}
\caption{Plot of upper kHz QPO frequency ($\nu_2$) vs. lower kHz QPO
frequency ($\nu_1$) for the Atoll and Z sources for which double kHz
QPOs have been published (see Table 1). Empty circles \new{correspond to 
Z sources, stars to Atoll sources} and filled circles to the
\object{Sco X-1} data (courtesy of M. van der Klis). The dotted line is the best
linear fit to the Atoll points, the dot-dashed line is the best linear
fit to the Z points \new{(excluding \object{Sco X-1}),}
and the thick line represents a fixed 3:2 ratio. 
\new{Errors of the frequencies
are typically around a few Hz.}
}
\label{figure1}
\end{figure}

\subsection{Correlation between frequencies}

Figure 1 shows a linear plot of the frequency of the upper kHz QPO
$\nu_2$ versus the frequency of the lower kHz QPO $\nu_1$. Stars
indicate Atoll sources, circles are for Z sources (filled circles for
\object{Sco X-1}). It is evident that a strong linear correlation exists between
the two frequencies, although slightly different for the two classes of
systems. A linear fit, $y = ax+b$, yields $a$=0.94$\pm$0.02,
$b$=350$\pm$15 for Atoll sources and $a$=0.85$\pm$0.01, $b$=383$\pm$8
for Z sources without \object{Sco X-1} (see dashed and dotted lines in Figure
1). Errors are at a 1$\sigma$ significance level. These relations, as
well as the $a$=0.73$\pm$0.01, $b$=469$\pm$7 
found for \object{Sco X-1} alone, are clearly not
compatible with the two frequencies having a constant 3:2 ratio (
$a$=1.5, $b$=0: continuous line in Fig. 1).

Following Abramowicz et al. (2003), we plot the histogram
of the $\nu_2 / \nu_1$ ratio distribution. 
A clear peak around 1.5,   as reported by
Abramowicz et al. (2003), can be seen (see Fig. 2). However,
this peak should be interpreted with some caution. The
distribution of the ratio of two quantities that follow a linear
correlation with $b\neq 0$ is largely determined by the distribution of
the observed values of one of the two variables. Consider that for a
linear relation

\begin{equation}
   y/x = a + b / x.
\end{equation}

\noindent  For high values of $x$ (i.e. $x > b$), 
which are the majority in Fig. 1,
the ratio will be asymptotically close to $a$, while for low values of
$x$ the ratio will be higher. To illustrate this, we simulated a sample
of frequency pairs around the correlation for atoll sources, but with
the lower kHz frequencies uniformly distributed in the 150-900
range. The corresponding histogram can be seen in the inset of Fig. 2:
\new{a  rapid increase toward low values of the ratio}
%$\nu_2 / \nu_1$=1.4 
is clearly visible.
This \new{increase} is the combined effect of the fact that the two
frequencies are correlated and of the range of observed frequencies.

\begin{figure}[ht]
\resizebox{\hsize}{!}{\includegraphics{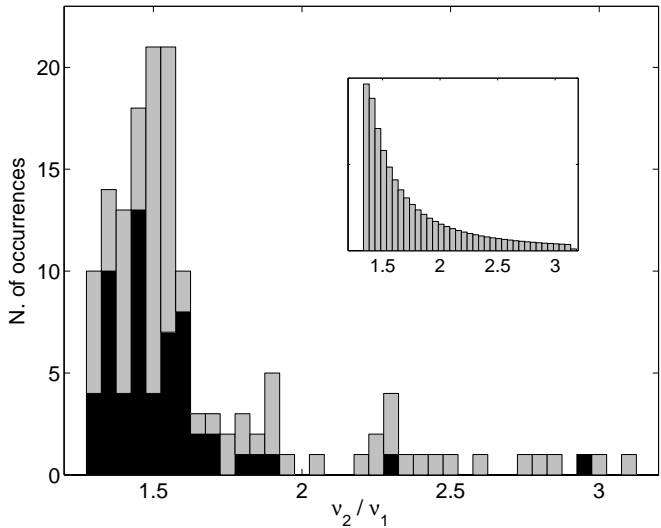}}
\caption{Distribution of the $\nu_2 / \nu_1$ values for points in Fig.
1. Black corresponds to Atoll sources, gray to Z sources (including 
\object{Sco X-1}). The inset shows the histogram corresponding to a simulated
uniform distribution of points around the correlation for atoll sources
(see text).} \label{figure2} \end{figure}

\subsection{A preferred 3:2 ratio in \object{Sco X-1}?}

While the effect described in the previous paragraph can (practically)
account for the observed peak in Fig.~2, it is still possible that the 
points in Fig.~1 intrinsically cluster around the intersection with the
3:2 line, even though their own correlation significantly differs from
that. This cannot be tested for most of the sources in Fig. 1, as
Abramowicz et al. (2003) point out, as the real distribution in
occurrence of the QPO pairs in these sources is hidden by the typical
selection procedures applied by the respective authors. It can however
be tested from the published \object{Sco X-1} data, which have been selected in
an unbiased way (see Abramowicz et al. 2003).

First, we produced a histogram of the distribution of one of the
frequencies, in this case $\nu_2$. This histogram can be seen in Fig. 3
and shows a broad excess around 900-950 Hz. A fit of the distribution
in Fig. 3 with a Gaussian model yields a centroid frequency of
$\nu_2^c$=930$\pm$8 Hz (1$\sigma$ error). Then, we fitted the $\nu_1$
vs. $\nu_2$ relation (the inverse relation of that shown in Fig. 1)
with a linear model, obtaining $a$=1.363$\pm$0.019 and $b$=-635$\pm$18.
Using this relation, we can assign a corresponding $\nu_1$ frequency to
$\nu_2^c$ and compute their ratio. The
resulting ratio is $R$=1.47$\pm$0.04. Notice however that a
Kolmogorov-Smirnov test shows that the hypothesis that the $\nu_2$
distribution is constant can only be rejected at a $2.4\sigma$
confidence level.

\begin{figure}[ht]
\resizebox{\hsize}{!}{\includegraphics{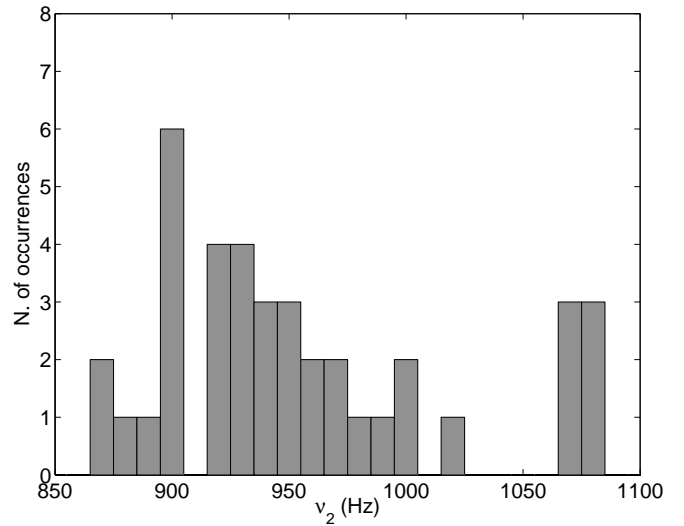}}
\caption{Distribution of the frequency of the upper kHz QPO $\nu_2$ for
the \object{Sco X-1} data analyzed by Abramowicz et al. (2003).} \label{figure3}
\end{figure}

There is therefore some marginal evidence for a clustering around a
frequency corresponding to a ratio of 1.5, marginal because of the
small number of points in this dataset. We note, however, that 
although we compared the distribution of $\nu_2$ against a uniform
distribution, there is no reason to expect a constant distribution in
QPO frequency. Also notice that in \object{Sco X-1} 
twin kHz QPOs are observed only
when the source is in the upper Normal Branch in the color-color
diagram (van der Klis et al. 1996) and therefore, by selecting
only data with two peaks, we introduce a bias toward low
frequencies (since QPO frequencies correlate with the position of
the source in a color-color diagram). Indeed, van der Klis et al.
(1996) report single-peak detections above 1090 Hz. In a number of
sources, a second low-frequency kHz QPO is found when the Power Density
Spectra are shifted according to the high-frequency QPO and then
summed, indicating that the second QPO might always be present,
although at a lower rms.

In order to investigate the actual distribution of kHz QPO frequencies
in bright LMXB we examined the high-frequency QPO detections of four
additional sources, as well as from  an expanded set of data from 
\object{Sco X-1}.

\section{Data analysis}

The selected data and the procedures used are described below.

\begin{figure}[ht] \resizebox{\hsize}{!}{\includegraphics{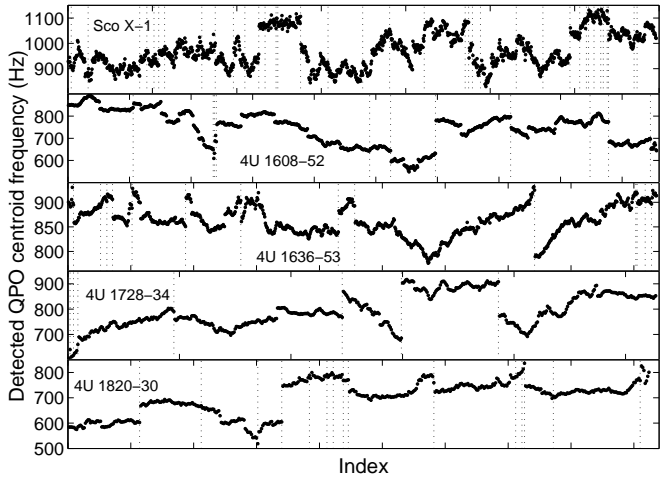}}
\caption{The series of measured QPO centroid frequencies for the five
sources analyzed. The X axis is 
\new{simply the PDS number, i.e. the frequencies
are plotted in sequence.}
All discrete
jumps in frequency correspond to gaps in the data. Gaps longer than one
hour are marked with dotted lines. Typical errors on the frequencies
are $1-5$ Hz.} \label{figure4} \end{figure}

\begin{itemize}

\item {\it \object{Sco X-1}:} we used the data analyzed by M\'endez \& van der
Klis (2000). Power density spectra (PDS) were accumulated from data
stretches 16 seconds long, then 8 consecutive PDS were averaged and
searched for high-frequency ($>$ 250 Hz) peaks. If two peaks were
found, only the highest one was considered. In all cases, the upper kHz
QPO was detected (see M\'endez \& van der Klis for more details). The
lower kHz peak was then recovered via a shift-and-add technique using
narrow (5 Hz) frequency bins: this procedure allows one to obtain an
averaged $\nu_2$ vs. $\nu_1$ relation (see Fig. 1 in M\'endez \& van
der Klis 2000, also reported by Miller 2003). The output of these
procedures are therefore a time series of $\nu_2$ frequencies and a
general correlation $\nu_2$ vs. $\nu_1$. We fitted this correlation
with a linear model, obtaining $a$=0.786$\pm$0.002, $b$=433$\pm$2.

\item {\it \object{4U 1608--52} and \object{4U 1728--34}:} 
we used the data analyzed by
M\'endez et al. (2001). The extraction procedure was
very similar to that for Sco X-1, with the following differences: PDS
were accumulated from 64s data stretches, a variable number of
consecutive PDS were averaged (in all cases less than 20) in order to
obtain a significant detection, and the {\it lower} kHz QPO peak was
selected, obtaining the $\nu_2$ vs. $\nu_1$ relation through shift and
add. A linear fit gives $a$=0.810$\pm$0.017, $b$=418$\pm$10 for 
\object{4U 1608--52} and $a$=0.897$\pm$0.024, $b$=420$\pm$17 for 
\object{4U 1728--34}.

\item {\it \object{4U 1636--53}:} we used the data analyzed by Di Salvo
et al. (2003). The extraction procedure was the
same as for \object{4U 1608--52} and \object{4U 1728--34}. To derive [$a$,$b$], 
we also used the points
by Jonker et al. (2002b). The linear fit for this
source gives $a$=0.673$\pm$0.017, $b$=540$\pm$14.

\item {\it \object{4U 1820--30}:} 
The extraction procedure was the same as for
\object{4U 1608--52} and \object{4U 1728--34}. 
Preliminary results were shown in M\'endez
(2002). We used all RXTE/PCA data from 1996 October 15th to 1999 March
1st.  For this source,  we obtained a fit with $a$=0.741$\pm$0.039,
$b$=466$\pm$29.

\end{itemize}

\begin{table*}
\begin{center}
\caption{\small Parameters of the linear fits and centroids of the
Gaussian distributions fitted to the histograms in Fig. 5 and
corresponding frequency ratios (see text). All errors are 1$\sigma$.}
\begin{tabular}{lcccccccc}
\hline \hline
Source           &  $a$            & $b$            & $\nu_\alpha$   & $\nu_\beta$     & $\nu_\gamma$     & $R_\alpha$           & $R_\beta$ &
$R_\gamma$ \\
 name            &                 &                & (Hz)           &  (Hz)           &   (Hz)          &                       &           &\\
\hline

{\rm Sco X-1}    & 0.786$\pm$0.002 & 432.5$\pm$ 1.5 &936 $\pm$ 2 & 1025$\pm$ 3 & 1077$\pm$3  & 1.46 $\pm$ 0.01 & 1.36 $\pm$ 0.01 & 1.32$\pm$0.01\\
{\rm 4U 1608-52} & 0.810$\pm$0.017 & 418.3$\pm$10.3 &594 $\pm$ 4 &  668$\pm$ 2 &  787$\pm$2  & 1.51 $\pm$ 0.02 & 1.44 $\pm$ 0.02 & 1.34$\pm$0.02\\
{\rm 4U 1636-53} & 0.673$\pm$0.017 & 539.8$\pm$14.1 &809 $\pm$10 &  856$\pm$ 2 &  897$\pm$3  & 1.34 $\pm$ 0.03 & 1.30 $\pm$ 0.02 & 1.27$\pm$0.02\\
{\rm 4U 1728-34} & 0.897$\pm$0.024 & 419.7$\pm$16.8 &756 $\pm$ 2 &  874$\pm$ 3 &             & 1.45 $\pm$ 0.03 & 1.38 $\pm$ 0.03 &\\
{\rm 4U 1820-30} & 0.741$\pm$0.039 & 466.2$\pm$29.0 &601 $\pm$ 1 &  731$\pm$ 2 &             & 1.52 $\pm$ 0.06 & 1.38 $\pm$ 0.06 &\\

\hline
\end{tabular}
\end{center}
\end{table*}

The resulting frequencies can be seen in Fig. 4, with vertical markers
indicating gaps longer than one hour.

For each source, we repeated the analysis described in Section 2.2. The
distributions of $\nu_2$ (for \object{Sco X-1}) and $\nu_1$ (for the other
sources) can be seen in Fig. 5. As all of the QPO frequency distributions
seem to be multi-peaked,
we fitted them with a sum of Gaussians. Two or three Gaussians were
needed to fit the observed peaks. The resulting centroids $\nu_\alpha$,
$\nu_\beta$ and $\nu_\gamma$ are reported in Table 2. The 
multi-Gaussian models used are also plotted in Fig. 5.

For each source we then used the best-fit $\nu_2 - \nu_1$ relation to
compute
the frequency ratios $R_\alpha$, $R_\beta$ and $R_\gamma$ corresponding
to the Gaussian peaks, with corresponding errors. The results are shown
in Tab. 2 and will be discussed in Sect. 4.  Notice that the errors on
$R_\alpha$, $R_\beta$ and $R_\gamma$ are dominated by the propagated
errors on the coefficients $a$ and $b$ from the linear fits to the
$\nu_2 - \nu_1$ relations. The errors on the ratios are therefore
correlated, i.e. even though the signs of the errors on the ratios are
not known, they are most likely the same for all ratios of each
individual source.

\begin{figure}[ht]
\resizebox{\hsize}{!}{\includegraphics{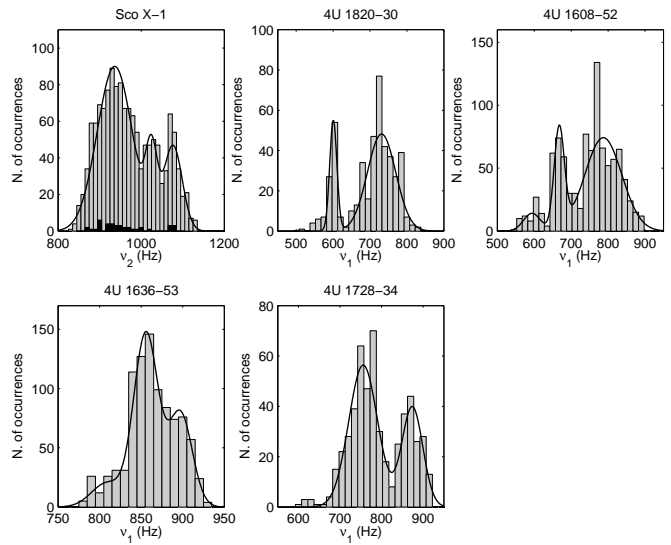}}
\caption{Distribution of the frequency of the upper kHz QPO $\nu_2$ for
\object{Sco X-1}, and of the lower kHz QPO $\nu_1$ for the other sources in our
sample (see text). \new{For \object{Sco X-1}, the black histogram represents the
frequency distribution of the upper kHz QPO 
in the data by Abramowicz et al (2003).}} 
\label{figure5} \end{figure}

\begin{figure}[ht] 

\begin{tabular}{cc}
\resizebox{4cm}{!}{\includegraphics{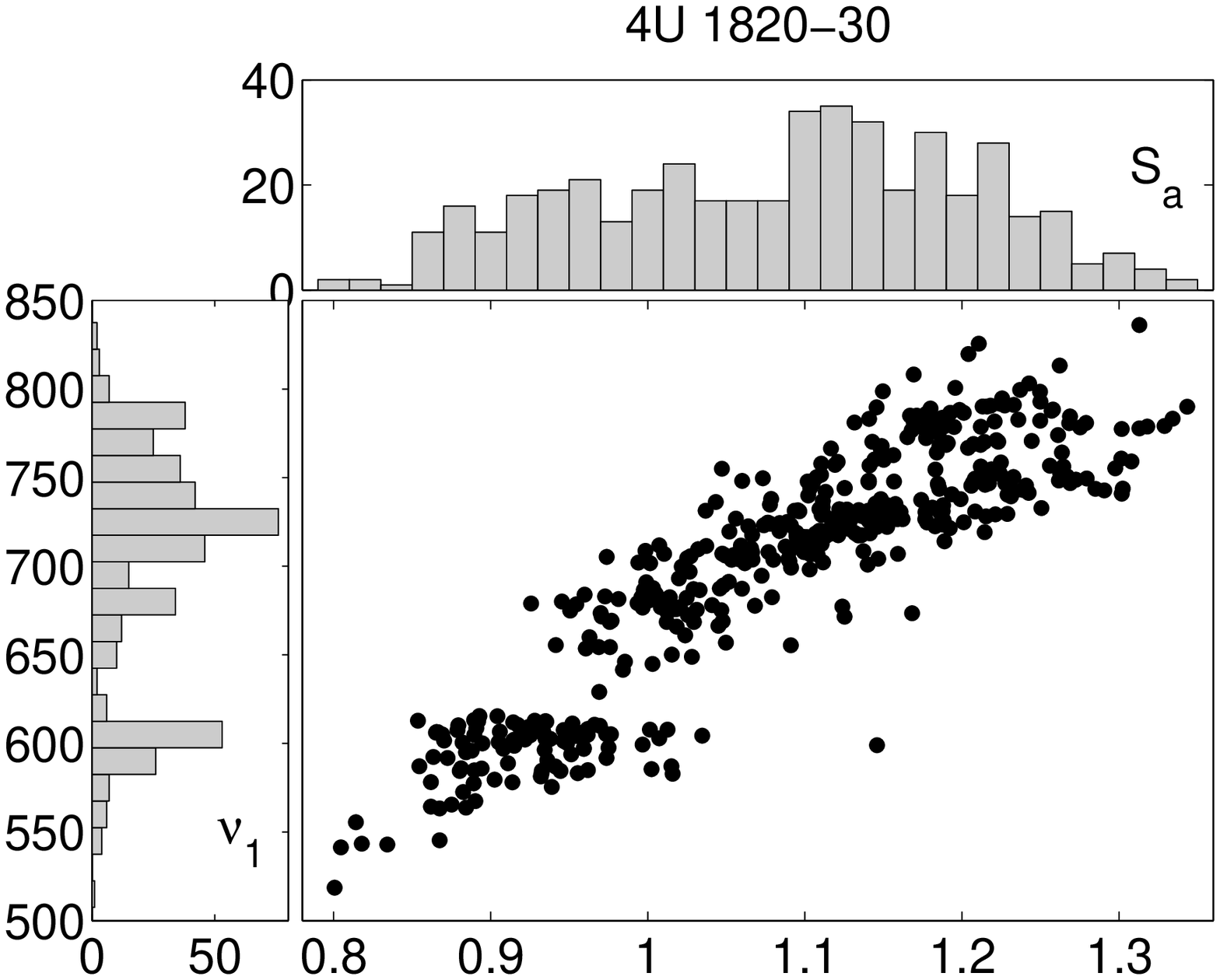}} &    %6cm for referee
\resizebox{4cm}{!}{\includegraphics{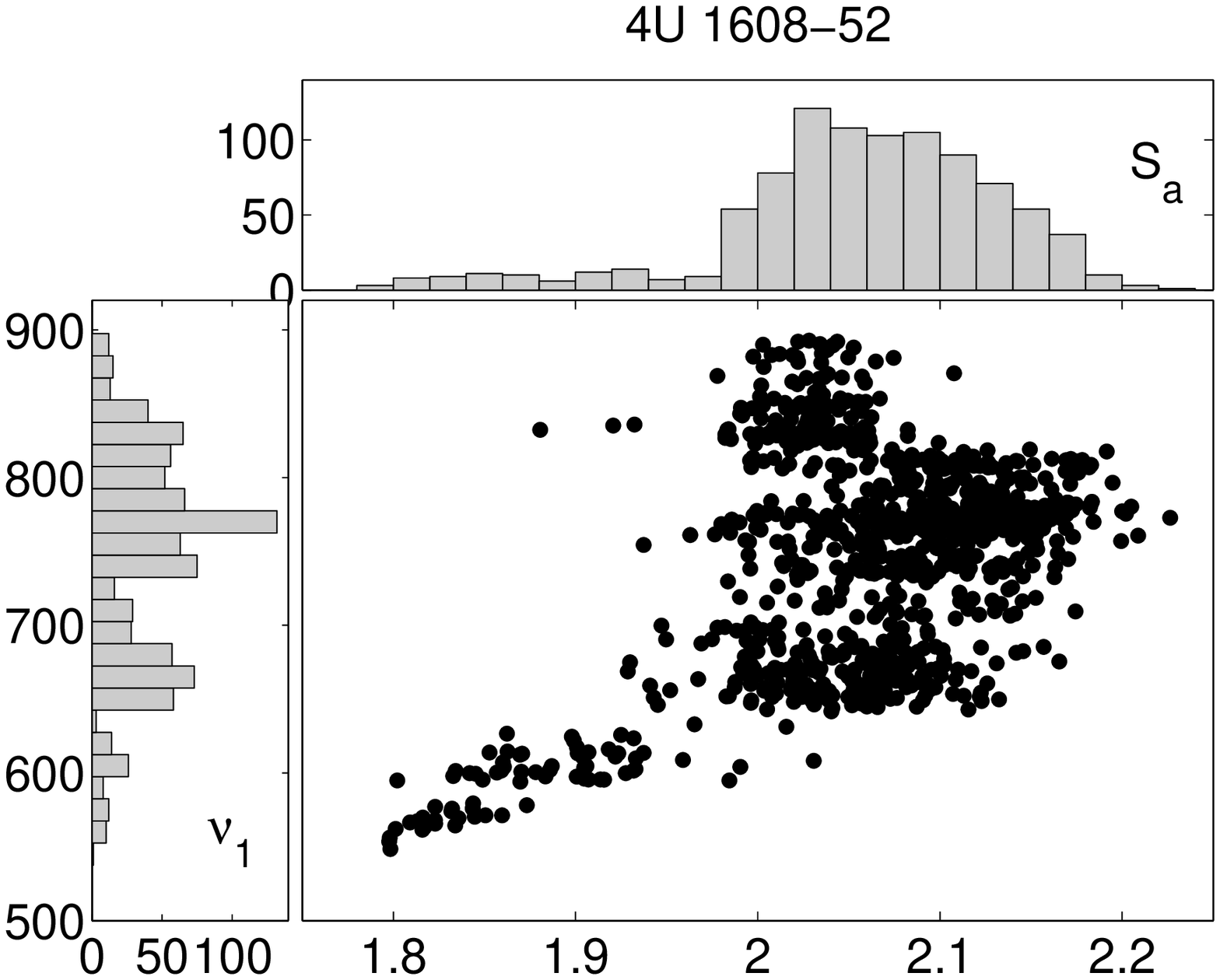}}\\
\resizebox{4cm}{!}{\includegraphics{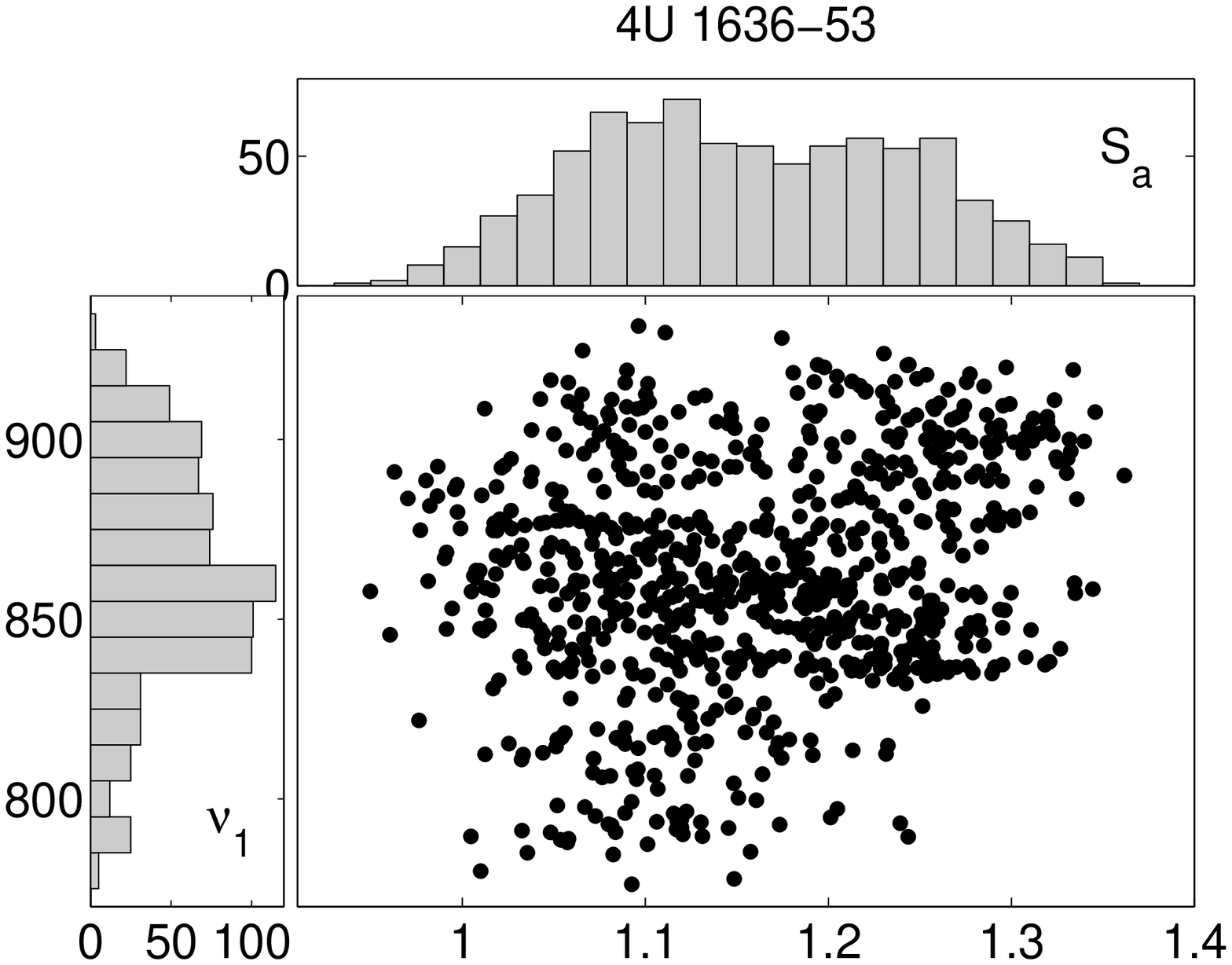}} &
\resizebox{4cm}{!}{\includegraphics{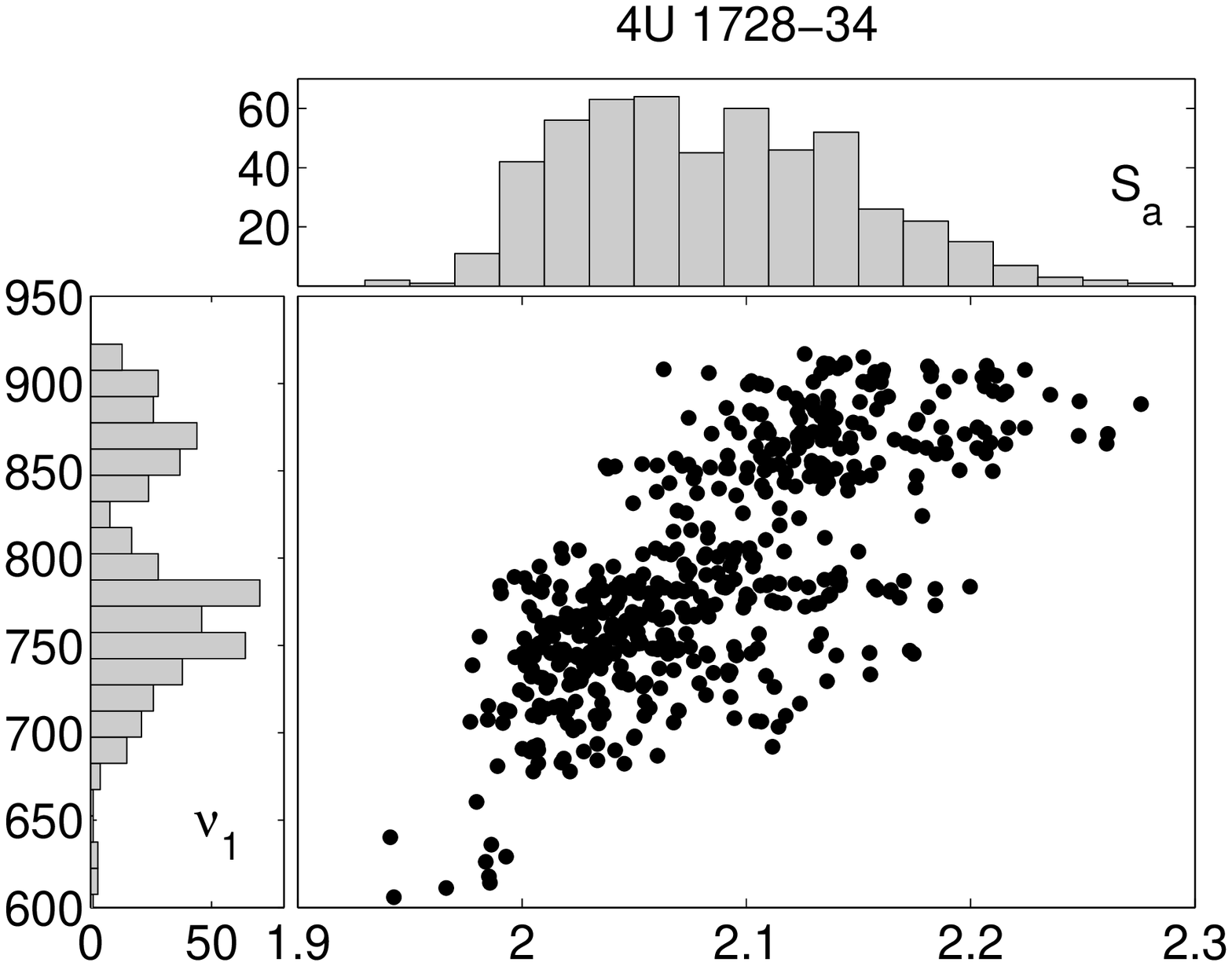}}\\
\end{tabular}
\caption{Frequency of the lower kHz QPO $\nu_1$ plotted versus the
position of the source along the color-color diagram, parametrized by
$S_a$ (see text), for the four atoll sources in our sample. The
histograms represent the marginal distributions of the single
parameters. Notice that the $\nu$ histograms are not identical to those
in Fig. 5, because not for all points we could extract a color.
} \label{figure6}
\end{figure}

As the frequency of QPOs in these systems is known to correlate with
the position on the color-color diagram (see e.g. Di Salvo et al.
2003),  it is possible that the non-uniformity of the distributions
shown in Fig. \ref{figure5} is the result of an analogous non-uniform
distribution of points along the color-color diagram of the source. In
order to check for such an effect for each of the sources, except 
\object{Sco X-1},
we produced a color-color diagram with each point corresponding to
one of our QPO detections, using the procedures described in Di Salvo
et al. (2003). In the case of \object{Sco X-1}, this proved not to be feasible
as the strong dead-time effects caused by the  high count rate and the
different offset pointings to this source prevented the production of a
single homogeneous color-color track.  We parametrized the position on
the color-color track following Di Salvo et al. (2003) and obtained for
each point an $S_a$ value, \new{i.e. an index that characterizes the position
along the color-color track.} 
The distribution of the points in the $\nu -
S_a$ plane is shown in Fig. 6, together with the marginal distribution
along the two axes. It is clear from the figure that the multi-peaked
distributions of QPO frequency do not have a corresponding multi-peaked
distribution in $S_a$. 
\new{For instance, in the case of \object{4U 1820--30} there is
a gap in the distribution of QPO frequencies between $\nu_1 \sim 600$
and $\nu_1 \sim 650$ Hz, but there is no gap in the $S_a$ distribution
at the corresponding values between $S_a \sim 0.9$ and $S_a \sim 1.0$} % end new
(see upper left panel in Figure \ref{figure6}).

%=====================================================================
% Here random walk hypothesis
%=====================================================================
 
\begin{figure}[ht]
\resizebox{\hsize}{!}{\includegraphics{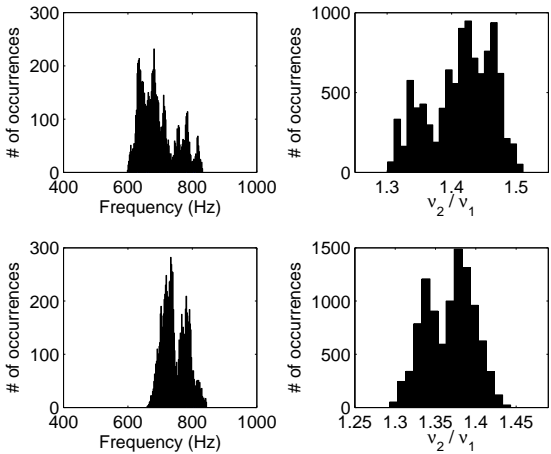}}
\caption{Left panels: Distribution of the frequencies of two random walks in
frequency (see text). Right panels: corresponding distribution of frequency
ratios assuming the $\nu_1 - \nu_2$ relation for \object{Sco X-1} (see Table 2).
} \label{figure7} \end{figure}

\new{
Having
ascertained that the distribution of QPO frequencies is not constant,
but shows multiple peaks, we need to establish whether these
peaks are simply the result of the random walk of the QPO frequencies.
It is known that the frequencies of kHz QPOs do not jump in frequency,
but show a sort of random walk in time (see Fig. \ref{figure4}).
In fact, this behavior is not a simple random walk, as
frequencies are only observed within a certain range. In order to
test whether such a random movement of frequencies in time could
produce peaks in the frequency distribution, we simulated a simple
random walk in frequency. We produced a series of 10000 frequencies
starting at 700 Hz and adding a
random value between $-6$ an $+6$ Hz to the previous value in the
sequence. The resulting histograms for two
realization of the random walk are shown in Fig. \ref{figure7}.
Adopting a linear relation between $\nu_1$ and $\nu_2$ as the one
shown in Table 2 for \object{Sco X-1}, 
for each simulation we also produced the distribution of $\nu_2/\nu_1$.
This very simplified model, which does not include a ``re-call'' force
to keep the frequency between fixed bounds, can clearly produce multiple
peaks in the frequency distribution, which for the reasons outlined above 
results in a sharp peak in the distribution of ratios.}  % end of new

%=====================================================================

\section{Discussion}

We investigated the possibility of a 3:2 resonance in the frequencies
of the kHz QPOs in bright accreting LMXBs from all available data in
the literature. We found that the strong linear correlation observed
between the lower and upper kHz QPO frequencies both in Atoll and Z
sources is {\it not} compatible with a single constant ratio, which can
be excluded with high significance (see Fig. 1). This conclusion is
also evident from Rebusco (2004), who suggests an ad hoc modification
to the resonance model in order to explain the observed correlation. We
showed that  even a constant distribution of the frequency of one of
the QPOs, together with an observed correlation between the two QPO
frequencies,  would result in a peaked histogram of frequency ratios. 
\new{Since resonance models of the QPOs predict that the frequency
ratios as well as the frequencies themselves should cluster around
specific values, and since the distribution of QPO frequencies does not
suffer from the bias of the distribution of frequency ratios,
we used the same \object{Sco X-1} data analyzed by Abramowicz et al. (2003) to
check whether there is evidence for a clustering around a particular
QPO frequency.} We found marginal evidence for an excess in the
distribution at a frequency consistent with a 3:2 ratio, whose
significance is difficult to assess given the scarcity of available
data, and whose nature might be the result of observational biases. 
\new{We
conclude that although the significance of the narrow peak 
in the distribution of QPO frequency ratios around a value of 3/2
reported by Abramowicz et al. (2003) in \object{Sco X-1} is
formally correct,
since the distribution of frequency ratios is biased, their result has no power
in testing the predictions of their resonance models; the peak in the
distribution of ratios that they find is probably due
to the distribution of QPO frequencies in \object{Sco X-1} being
mostly above the intersection with the line of a constant 3:2 ratio.}

The analysis of a sample of high-frequency oscillations in five systems
(including a more extensive coverage of \object{Sco X-1}) shows that the
$\nu_2-\nu_1$ relation is rather similar between different systems, and
that the distribution of the observed frequencies is not flat in the
observed range, but shows two or three peaks, depending on the source.
Abramowicz et al. (2003) reported the presence of a second
peak, although at low significance, corresponding to the six points
above $\nu_2$=1050 Hz in Figs. 1 and 3. From Fig. 5, it is evident that those
points correspond to our high frequency component $\nu_\gamma$. 
Using
the measured $\nu_1-\nu_2$ relations, it is possible to associate a
frequency ratio to the centroids of these peaks. The lowest peak
corresponds to a ratio close to 1.5 in four systems, while for 
\object{4U 1636--34} it does not.

There is no a priori reason for such a distribution to have
a flat shape, since it is known that the fractional rms of these
oscillations decreases at high and at low frequencies (see M\'endez
et al. 2001 and Di Salvo et al. 2003).
However, the frequency dependence of fractional rms does not show any
evidence of being multi-peaked, and its maximum does not coincide
with any of the maxima of the distributions in Fig. 5.

We have shown that the multi-peaked nature of the distribution of QPO
frequencies is not due to a combination of the source spending more
time on certain positions in the color-color diagram, and of a
correlation between position on the color-color diagram and QPO
frequency (Fig. 6). The lack of correspondence of the multi-peaked
distribution of QPO frequency with the distribution of the $S_{a}$
values contradicts scenarios in which the QPO frequency and the
source spectrum are both determined only by the radius of the inner edge of
the disk (van der Klis 2001). Notice that in particular \object{4U 1820--30},
while showing a good correlation between $\nu_1$ and $S_a$, clearly
features a gap in the distribution of $\nu_1$ not associated to a
corresponding gap in $S_a$.

 From the few instances of double high-frequency QPOs in black-hole
systems there is some evidence for a preferred fixed ratio. The data
are sparse and rather stable frequencies have been observed in these
pairs.  In the case of BH binaries the preferred ratio can
also be the result of sampling, i.e. the few detections available to
date could come from the upper region of a correlation similar to that
in Fig. 1, in which case the observed ratio would \new{not have a simple
physical interpretation in terms of a resonance mechanism.} For
neutron-star systems, where a large number of detections of pairs with
variable frequencies is available, a single fixed ratio can be
excluded.  The presence of an additional external force as in the model
by Rebusco (2004) (see also Abramowicz et al. 2003b)
could explain the observed deviation from a fixed 3:2
ratio, at the expense of a considerable complication of the model, with
the addition of an unknown parameter which biases the observed values
away from a direct measurement of a 3:2 ratio. However, the presence of
multiple peaks as those shown in Fig. \ref{figure5} would complicate
the interpretation further, as the presence of other preferred
frequencies (or ratios) would also have to be explained.

If the kHz QPOs are due to a resonance mechanism in the accretion disk,
the multi-peaked distribution of QPO frequencies (Fig. 5) suggests that
not just the 3/2, but more than one resonance is at work. Using the
$\nu_2 - \nu_1$ relations in Sect. 3, we can assign a frequency ratio
to each of the peaks of the QPO distributions in Fig. 5. The inferred
ratios are close to 3/2, 7/5, and 4/3 in the cases of \object{Sco X-1} 
and \object{4U 1608--52}, 4/3, 9/7 and 5/4 in the case of 
\object{4U 1636--53}, and 3/2 and 7/5
in \object{4U 1728--34} and \object{4U 1820--30}. However, the errors are
sufficiently large that more than one identification is possible. 

 From the ratios shown in Table 2, it is apparent that \object{4U 1636--53} is
different from the other sources in that the frequency ratios in this
source are systematically lower than in the other four sources. While
for all other sources the largest ratio is close to 1.5, the largest
ratio in \object{4U 1636--53} is $1.34 \pm 0.03$, significantly smaller than
1.5. From the sample of sources in Table 2, \object{4U 1636--53} is the only one
that during the RXTE observations used for our analysis appeared always
in a region of the color-color diagram characteristic of sources at
high inferred mass accretion rates, and the QPOs appeared always at
high frequencies (Di Salvo et al. 2003). Assuming there is a relation
between \new{QPO frequency} and radius of the inner edge of the disk
(e.g., Miller et al. 1998), this suggests that during these
observations the inner edge of the disk in \object{4U 1636--53} was always
relatively close to the neutron star. On the other hand, based on their
intensity, color-color diagrams, and range of QPO frequencies, the
other sources appeared to cover a larger range of mass accretion rate,
which suggests that in those sources the inner radius of the disk
covered a larger range than in \object{4U 1636--53}.

\new{
It has already been shown (Stella \& Vietri 1999) that the
identification of the QPOs with the azimuthal and the periastron
precession frequencies, $\nu_{\theta}$ and $\nu_{\rm p} = \nu_{\theta}
- \nu_{\rm r}$, respectively, reproduces qualitatively the QPO
frequency-frequency correlations. 
[We note in passing that the alternative identification of
the QPO frequencies with the vertical and radial epicyclic frequencies
(e.g., Kluzniak et al. 2004) implies frequency-frequency correlations
in contradiction with the observed data. For instance, unless another
mechanism is at work, this interpretation predicts that the frequency
difference between the kHz QPOs, $\nu_2 - \nu_1$, increases as the
frequencies increase, opposite to what is observed in several sources; see
e.g., M\'endez et al. (1998a)]. 
The $\nu_\theta$,$\nu_p$ identification can also account
for the range of QPO frequency ratios seen in \object{4U 1636--53}, compared to
those seen in the other sources. The ratio $R = \nu_{\theta} / \nu_{\rm
p}$ increases monotonically with $r$, and hence decreases as the
frequencies increase (notice that, on the contrary, the ratio
$\nu_{\theta} / \nu_{\rm r}$ increases toward higher frequencies). 
Since in our observations of \object{4U 1636--53} the kHz
QPOs never reached values as low as those seen in the other
sources, the frequency ratio did not reach high values either. 
We expect that when the mass accretion rate in \object{4U 1636--53}
decreases, and the kHz QPO move to lower frequencies, the ratio of QPO
frequencies will increase and cluster around 1.5 as in the other
sources.
}

\new{
There is in principle no a priori
reason why higher frequency ratios such as 2/1, 3/1, 4/1, etc. 
should not appear. 
%(The
%parametric resonance model proposed by Abramowicz et al. 2003, and
%Rebusco 2004, cannot explain such large ratios; Kluzniak 2004, private
%communication.) 
With the identification of the observed frequencies with those of
Stella \& Vietri (1999), 
for a neutron star mass in the range $1.4
- 2.2 \msun$, a frequency ratio $R \geq 2/1$ would only occur when the
frequency of the lower kHz QPO is $\simless 300 - 500$ Hz, while a
ratio $R \geq 3/1$ requires the frequency of the lower kHz QPO to be
$\simless 150 - 200$ Hz. In none of the observations presented in this
paper did the lower kHz QPO frequency reach such a low value.
However, there are some reports in the literature of
lower kHz QPO being detected in this frequency range. For instance, the
lowest frequencies reached by the kHz QPO in the atoll source 
\object{4U 0614+09} are 
$\nu_1 = 153.4 \pm 5.6$ and $\nu_2 = 449.4 \pm 19.5$ (van
Straaten et al. 2000), such that $R = 2.9 \pm 0.2$. In the Z source 
\object{GX 5--1}, the lowest frequencies so far observed for 
a pair of kHz QPOs are
$\nu_1 = 156 \pm 23$ Hz and $\nu_2 = 478 \pm 15$ Hz (Jonker et al.
2000), for which $R = 3.1 \pm 0.4$.
}

%--------------------------------------
% random walk part in the discussion
%--------------------------------------

\new{
Finally, we showed that a simple frequency random walk could in
principle reproduce the observed multi-peaked frequency distributions as
well as the distribution of frequency ratios. Although we have used a
very simple approach to simulate the drift of QPO frequency with time (e.g.
the real frequencies do not follow a pure random walk, the observations
sample many separate intervals of the real frequency evolution, etc.),
it is clear that a multi-peaked distribution of QPO frequencies (or
frequency ratios) could also result from purely random processes.  A
possible test of the idea that the observed QPO-frequency distribution
in these neutron star systems is indeed produced by an underlying
resonance mechanism would be to accumulate new (independent) samples of
frequencies of similar size to the ones presented here, and check
whether the distribution of QPO frequencies in these new samples show
peaks with centroid frequencies that are consistent with those observed
here (although some peaks could be missing while some new peaks may
appear, depending on the range of frequencies sampled by the sources
during the new observations). If this is not observed, we would have to
conclude that the peaks are due to the random motion of the frequency.
Since at present we cannot discard this possibility, our results weaken
the support for resonance models of the kHz QPOs in neutron stars. 
} % end of new

%---------------------------------------------------------------------

\section{Summary and conclusions}

\new{
We showed that:
}

\begin{itemize}

\item \new{The relation between the two simultaneous kHz QPO frequencies in
a large sample of sources, including \object{Sco X-1}, is significantly
different from a $\nu_2 = 1.5 \nu_1$ relation.}

\item \new{Since in all these sources the two QPO frequencies are linearly
correlated, the distribution of the ratios of the QPO frequencies would
be peaked even if individually the frequencies are distributed more or
less uniformly. Since resonance models predict that {\em both} the
ratio of frequencies as well as the frequencies themselves should
cluster around specific values, the distribution of frequency, and not
of frequency ratios, should be used to test the predictions of those
models.}

\item \new{The hypothesis that in the same \object{Sco X-1} 
data used by Abramowicz
et al. (2003) the distribution of QPO frequencies is significantly
different from constant can only be rejected at a $2.4\sigma$
confidence level.}

\item \new{The distribution of QPO frequencies in \object{Sco X-1}, 
\object{4U 1608--52}, \object{4U 1636--53}, \object{4U 1728--34}, 
and \object{4U 1820--30} is multi-peaked, with the peaks
occurring at frequencies compatible with $\nu_2 / \nu_1$ ratios of 
3/2, 4/3, 5/4, 7/5, and 9/7, not all ratios appearing in all sources.
This is considerably different from the claim by Abramowicz et al.
(2003) that the frequency ratio for Sco X-1 shows a narrow peak at 1.5,
which we showed in Sect. 2 not to be statistically founded.
}

\item \new{The multi-peaked nature of the frequency distribution appears
difficult to explain by parametric resonance models, since the
perturbation mechanism that drives the ratio away from the main, 3/2,
resonance would have to be able, at the same time, to produce the other
resonances as well.}

\item \new{A random walk of the QPO frequencies qualitatively reproduces the
observed QPO frequency distribution as well as the frequency ratio
distribution.
This means that, until the presence of the same peaks at the same locations
is confirmed by (large) independent datasets, it is premature to speculate
further about the presence of preferred ratios.}

\end{itemize}

\begin{acknowledgements}

We thank Michiel van der Klis for providing the \object{Sco X-1}
frequencies used in Abramowicz et al. (2003), and Gabriele
Ghisellini, Sergio Campana, Rudy Wijnands, Michiel van der Klis, Peter
Jonker and Tom Maccarone for useful discussions.

\end{acknowledgements}

\end{document}